# Fractional Dynamical Model for the Generation of ECG like Signals from Filtered Coupled Van-der Pol Oscillators

Saptarshi Das* and Koushik Maharatna

*School of Electronics and Computer Science,*

*University of Southampton, Southampton SO17 1BJ, United Kingdom.*

**Authors' Emails:**

sd2a11@ecs.soton.ac.uk, s.das@soton.ac.uk (S. Das*)

km3@ecs.soton.ac.uk (K. Maharatna)

**Corresponding author's phone number:** +44(0)7448572598

**Abstract**

In this paper, an incommensurate fractional order (FO) model has been proposed to generate ECG like waveforms. Earlier investigation of ECG like waveform generation is based on two identical Van-der Pol (VdP) family of oscillators which are coupled by time delays and gains. In this paper, we suitably modify the three state equations corresponding to the nonlinear cross-product of states, time delay coupling of the two oscillators and low-pass filtering, using the concept of fractional derivatives. Our results show that a wide variety of ECG like waveforms can be simulated from the proposed generalized models, characterizing heart conditions under different physiological conditions. Such generalization of the modelling of ECG waveforms may be useful to understand the physiological process behind ECG signal generation in normal and abnormal heart conditions. Along with the proposed FO models, an optimization based approach is also presented to estimate the VdP oscillator parameters for representing a realistic ECG like signal.

**Keywords:** Electrocardiogram (ECG); fractional calculus; delay differential equation (DDE); phase space; Van-der Pol (VdP) oscillator

## 1. Introduction

Mathematical modelling of biological signals is quite challenging and is an emerging field of research. This provides better understanding of the underlying physical phenomena which results in different physiological signals in human body like Electrocardiogram (ECG), Electroencephalogram (EEG), Electromiogram (EMG) etc. [1]. These physiological signals are widely used, over the years by the clinicians to diagnose irregular behaviour of human organs. Therefore, mathematical modelling for the generation of such signals from a system theoretic point of view may lead us to the root of the complicated physiological processes, responsible for their generation, to detect healthy and unhealthy behaviour of human organs.

In this paper, a class of new fractional dynamical models are proposed in order to generalize various healthy and diseased ECG waves, in particular the QRS-complex.

The fundamental principle behind generation of any complex biological waveform is governed by few set of nonlinear differential equations. Several dynamical system theoretic approaches have been proposed to mathematically model ECG waveform and the action potentials, generated in different nodes of human heart [2]. Various arguments are reported in the literature, for example by Babloyantz and Destexhe [3], about the fact whether the heart perfectly behaves like a periodic oscillator or not. It is arguable that in an actually recorded ECG signal, there might be slight morphological difference between the consecutive heartbeats (due to noise or other external stochastic disturbances like artefacts etc.) which produces chaos-like random wandering of the states in the phase space [3]. But there is no doubt that human physiological systems exhibit some kind of complex periodic waveforms like other different cardiovascular signals, e.g. respiration, ECG, heart rate variability (HRV), blood pressure, blood flow etc. [4], [5], [6], which motivates modelling of such systems using equivalent coupled oscillators. Representing each of the physiological oscillations with the corresponding characteristic frequency, a coupled oscillator model is developed in Stefanovska *et al.* [5], [6] for cardiovascular system in human body. Three ordinary differential equation based model was proposed by Zbilut *et al.* [7] to represent variation in the inter-beat length, although the obtained signal is not similar to the P-QRS-T complex of an ECG signal. The action potentials generated from sino-atrial (SA) and atrio-ventricular (AV) nodes have been modelled using mixed VdP/Duffing type relaxation oscillators by Grudzinski and Zebrowski [8]. Synchronization studies have been done between the SA and AV nodes represented as two coupled VdP oscillators with uni-directional and bi-directional couplings having external forcing which are expressed in the general form of a pair of Lienard equations by Santos *et al.* [9]. Other nonlinear state-space approaches have also been able to successfully generate the ECG waveform like in [10], [11], [12] while considering the characteristics of P, Q, R, S, T waves as a state variable.

Among different approaches, Kaplan *et al.* [13] has shown that a nice ECG like waveform can be generated by using a simple model of two coupled VdP family of oscillators. In the pioneering work [13], the authors have proposed a model having two identical filtered VdP oscillators and represented it by two coupled systems of delay differential equations (DDEs). Each of the coupled oscillators is having three state variables. The first equation is responsible for introducing the nonlinearity in the model due to having cross-product of different state variables. The second equation introduces the time delay coupling by mutual injection of delayed signals from two oscillators. The third state equation acts as a low-pass filter to stabilize the amplitude of the oscillator. The integer order model actually do not have the capability of generalizing different ECG waveforms, as the long term memory effect is not there unlike fractional differential equations. This motivates us for exploring similar fractional dynamical models for ECG signal generation. In this paper we have studied analogous FO dynamical models of the coupled VdP oscillator to represent various healthy and unhealthy ECG like signals. Such an approach leads to generalization of a wide variety of ECG like waveforms which may give us some clinically significant information about the health of human heart.

Fractional calculus is a 300 years old mathematical tool and has recently been popular in mathematical modelling of many real world systems [14]. The main concept of this particular branch is based on representing successive differentiation and integration of a function to take any arbitrary real value. The main advantage of fractional calculus based approach is that it is capable of incorporating long term memory behaviour of a system in the model, implying capturing the history or memory effect, in comparison with the classical integer order derivatives, describing the rate of change of a variable at a particular time instant. In the solution of fractional differential equations the memory kernel decays as a power law whereas the memory kernel becomes the Dirac Delta function for ordinary differential equations [14]. Thus, the fractional differential equations have the better ability of capturing the dynamics of natural systems with higher capability of remembering the past evolution of the function due to the infinite dimensional nature of fractional derivatives [15]. Due to having the higher capability of modelling real systems, fractional calculus based approaches pervaded several disciplines like control [16], signal processing [17], [18], nonlinear dynamical systems theory [19], physics [20], biology [21] etc. Magin in [22], [23] has shown several fractional dynamical models for biological systems like e.g. nerve excitation, membrane charging, one-dimensional cable model for nerve axon, vestibular ocular models, bio-electrode models, viscoelastic models of cells, tissues, respiratory mechanics etc. The fractional time derivative operators in the dynamical models refer to some sort of fractal nature or statistical self-similarity in the model which is quite common in time series signals obtained from biological systems [21]. Recent studies have been focussed on other biological systems modelling using fractional calculus like red blood cell mechanics [24], viscoelasticity of human brain tissue [25], cell rheological behaviour [26], human calcaneal fat pad [27], protein dynamics [28], pharmacokinetic drug uptake model [29], modelling HIV infection [30], [31], [32], tissue modelling [33], dielectrics in fresh fruits and vegetables [34], system identification approaches for muscle [35] etc. Hence, it is logical to study analogous fractional dynamical model for ECG like signal generation.

The focus of the present paper is to develop a generalized template for ECG signal generation using fractional order modelling technique. In this effort two novel classes of models using coupled VdP oscillators have been proposed to describe various ECG like waveforms under normal and abnormal conditions of the human heart. It is shown that among several characteristics of ECG waves, particularly the ventricular characteristics i.e. the QRS complex and also the variability in heart rate can suitably modelled using the proposed fractional dynamical model of coupled oscillator system. Among the proposed models, the first class consider incommensurate fractional order dynamics in two identical VdP oscillators which are coupled by equal time delays. Further improvement of the FO coupled VdP oscillator system has been done by considering different time delays among the two filtered FO oscillators which gives the second class amongst the proposed models. The parameters of this particular coupled oscillator structure are estimated next to mimic a real ECG waveform, using a global optimization framework.

Rest of the paper is organised as follows. Section 2 discusses the background and motivation of FO modelling of ECG signals. Section 3 proposes the FO coupled oscillator system and shows simulation studies with it to draw an analogy of its response with ECG waveforms. In section 4, time delay couplings are considered to be different for the identical VdP oscillators and the parameter estimation results for the DDEs are presented. The paper ends with the conclusions as section 5, followed by the references.

## 2. Background and motivation of fractional dynamics in modelling ECG waveforms

### 2.1. *Basics of fractional calculus and fractional order nonlinear dynamical systems*

FO systems are governed by FO differential equations where the derivative orders can take any arbitrary real value instead of the integer values in classical integer order differential equations [15]. The fractional derivatives in the fractional differential equations can be defined using any standard definitions of fractional differ-integral viz. the Riemann-Liouville, the Grunwald-Letnikov or the Caputo definition according to the nature of the system's behaviour [14]. A linear FO differential equation contains constant coefficients to each of the terms having FO time derivative and conversely its nonlinear counterpart will have multiplication or higher powers associated with the state variables ($x$) which can be viewed as a variable coefficient one. Any linear or nonlinear fractional differential equations can be represented in terms of standard state space notation $\dot{x} = f(x)$, by replacing the first order time derivative by analogous fractional differ-integral term with respect to time.

FO nonlinear dynamical systems [19] are generally represented by the following FO state space formulation

$$D^{q_i} x_i(t) = f_i(x_1(t), x_2(t), \cdots, x_n(t)); \quad 0 < q_i < 2, x_i \in \Re^n \qquad (1)$$
$$x_i(0) = c_i; \quad i = 1, 2, \cdots, n$$

Here, $D^{q_i}$ represents the time derivative of any arbitrary order $q_i$ and $c_i$ are the initial conditions of the individual state variables $x_i$. For a system with constant fractional orders associated with each state variable i.e. $q_1 = q_2 = \cdots = q_n$ is known as a commensurate order nonlinear fractional dynamical system. Conversely for a system where the fractional orders ($q_i$) are different in the system of ordinary fractional differential equations, is known as an incommensurate order fractional dynamical system. Both of these modelling philosophies have been applied in this paper to produce ECG like signals.

Just like the classical VdP oscillators its FO counterpart also shows undamped periodic waveform [36]. Fractional VdP oscillator with incommensurate order models has been studied in Tavazoei et al. [37]. Further studies have been done on FO damping effects on the VdP oscillator by Chen and Chen [38]; and with external forcing by Ge and Hsu [39]. In the present paper, amongst various definitions of fractional derivative, the Caputo definition (2) is used in order to describe each fractional order operator in (1).

$$D^q x(t) = I^{m-q} x^{(m)}(t), \quad q > 0 \qquad (2)$$

where $m = \lceil q \rceil$ is the smallest integer that is larger than $q$ and $x^{(m)}(t)$ is the integer order $m$-times successive differentiation of $x(t)$ with respect to time $t$. $I^\beta$ is the Riemann-Liouville integral operator of order $\beta > 0$ given by (3).

$$I^\beta f(t) = \frac{1}{\Gamma(\beta)} \int_0^t (t-\tau)^{\beta-1} f(\tau) d\tau \qquad (3)$$

The gamma function $\Gamma(\cdot)$ is given by the following equation

$$\Gamma(n) = \int_0^\infty t^{n-1} e^{-t} dt \qquad (4)$$

The Laplace transform of fractional differ-integral is given by (5).

$$\int_0^\infty e^{-st} D^\beta f(t) dt = s^\beta F(s) - \sum_{k=0}^{m-1} s^{\beta-k-1} D^k f(0) \tag{5}$$

where $F(s)$ is the Laplace transform of function $f(t)$ and $s$ being the complex frequency.

For the simulation of linear FO systems the rational approximation methods are often referred viz. Charef's method, Oustaloup's method etc. [14]. The rational approximation methods replace each FO differ-integral operators ($s^q$) by suitable higher order transfer functions which maintains a constant phase of $q\pi/2$ within a suitably chosen frequency band. Historically, the study of fractional nonlinear dynamical systems started with Charef's recursive approximation of FO time derivatives [40], later Tavazoei and Haeri [41] have shown that the nonlinear system might show fake chaos with such rational approximations. For numerical solution of FO nonlinear differential equations, the Adams-Bashforth-Moulton predictor-corrector method is widely used [42]. But such an algorithm cannot solve delay differential equations as has been used in the present paper. A recent modification of the predictor-corrector algorithm for solving FO delay differential equation has been proposed by Bhalekar and Gejji [43]. The early investigation of FO VdP oscillator was done using Charef's rational approximation technique in Barbosa *et al.* [44] which has slightly lower accuracy in phase of the frequency response than that with the Oustaloup's method. Petras in [19] has shown that an Oustaloup's recursive approximation (ORA) can reliably used for numerical simulation of fractional nonlinear systems, using the MATLAB based Toolbox *Ninteger* [45] which has been used in this paper. The ORA approximates a FO differ-integral operator ($s^\beta, \beta \in \Re, \beta \in [-1,1]$) with an equivalent analog filter given by (6).

$$s^\beta \simeq K \prod_{k=-N}^{N} \frac{s + \omega'_k}{s + \omega_k} \tag{6}$$

where the poles, zeros, and gain of the filter can be recursively evaluated as:

$$\omega_k = \omega_b \left(\frac{\omega_h}{\omega_b}\right)^{\frac{k+N+\frac{1}{2}(1+\beta)}{2N+1}}, \omega'_k = \omega_b \left(\frac{\omega_h}{\omega_b}\right)^{\frac{k+N+\frac{1}{2}(1-\beta)}{2N+1}}, K = \left(\frac{\omega_h}{\omega_b}\right)^{-\frac{\beta}{2}} \prod_{k=-N}^{N} \frac{\omega_k}{\omega'_k} \tag{7}$$

Thus, any signal $f(t)$ can be passed through the filter (6) and the output of the filter can be regarded as an approximation to the fractionally differentiated or integrated signal $D^\beta f(t)$. In (6)-(7), $\beta$ is the order of the differ-integration, $(2N+1)$ is the order of the filter and $(\omega_b, \omega_h)$ is the expected fitting range of frequency. In the present study in all cases, 5$^{th}$ order ORA has been adopted to represent the integro-differential operators within the frequency band of $\omega \in \{10^{-2}, 10^2\}$ rad/sec. The choice of ORA lower and upper cut-off frequencies and bandwidth can be justified in a sense that all ECG signal in various condition generally lie within this wide spectrum. Commonly in ECG signal processing literatures, a band-pass filtering is employed within 0.1 to 30 Hz to retain only the necessary information intact [1]. The ORA bandwidth has been chosen to be large enough to ensure that the informative frequency components of the desired signal lie in most flat phase region i.e. around the centre of the ORA approximation range. This also alleviates the risk of any possible loss of flatness in the ORA filter phase near the lower and higher cut-off edges, as studied by Das *et al.* [46].

## 2.2. Background of using filtered coupled Van-der Pol oscillators to produce ECG like signals

The oscillator model studied here is related to the VdP equation which was historically developed to model human heartbeat [47].

$$\dot{x} = y + \varepsilon(1 - \mu y^2)x, \quad \varepsilon > 0, \mu > 0$$
$$\dot{y} = -x$$
(8)

Here, the nonlinear cross-product term with coefficient $\varepsilon$ is responsible for the damping so as to stabilize the oscillator's amplitude and plays a significant role in shaping the limit cycle. It was shown by Kaplan *et al.* [13], [47] that the above VdP oscillator can be suitably modified using a half-wave rectification and low-pass filtering action as the third state variable to produce ECG like signals.

$$\dot{x} = y + \varepsilon(1 - \mu z)x$$
$$\dot{y} = -x$$
$$\dot{z} = \left[\left((|y| - y)/2\right) - z\right]/T$$
(9)

Here, $T$ is the time constant associated with the low-pass filtering action, $\varepsilon$ and $\mu$ are the control parameters and are responsible for the amplitude stabilization as well as shaping the limit cycle. The integer order coupled oscillator model in [13] considered two identical filtered VdP oscillators (10) which are connected to each other by their respective second state variables with a gain ($\alpha$) and time-delay ($\tau$).

$$\left.\begin{aligned}
\dot{x}_1 &= y_1 + \varepsilon(1 - \mu z_1)x_1 \\
\dot{y}_1 &= -x_1 + \alpha\left[x_2(t-\tau) - x_1(t-\tau)\right] \\
\dot{z}_1 &= \left[\left((|y_1| - y_1)/2\right) - z_1\right]/T
\end{aligned}\right\}$$
$$\left.\begin{aligned}
\dot{x}_2 &= y_2 + \varepsilon(1 - \mu z_2)x_2 \\
\dot{y}_2 &= -x_2 + \alpha\left[x_1(t-\tau) - x_2(t-\tau)\right] \\
\dot{z}_2 &= \left[\left((|y_2| - y_2)/2\right) - z_2\right]/T
\end{aligned}\right\}$$
(10)

In Kaplan *et al.* [13] typical parameters for the coupled oscillator are suggested as $T = 20, \alpha = 0.05, \varepsilon = 2, \mu = 1, \tau = 100$ but the initial conditions, required for the simulations are not suggested. We found by rigorous simulation study that the coupled oscillators (10) do not show sustained periodic oscillation unless the initial conditions for the two VdP systems are the same. In all the simulations we used a fourth order Runge-Kutta algorithm with fixed step size of 0.01 seconds to solve the initial value problems. Also, for the simulation studies reported in this paper, the initial values of the six state variables in equation (10) are chosen as some non-zero value e.g. $x_0 = \begin{bmatrix} 0 & 10^{-6} & 0 & 0 & 10^{-6} & 0 \end{bmatrix}$. It is also found that since the two oscillators in (10) are identical in nature, with the same initial condition, the corresponding states of each filtered VdP oscillator will evolve in the same manner. Since in the pioneering work [13], the time delays corresponding to each of the second state variable are considered to be same, the two delayed signals of each oscillator in (10) become identical. As a result, the coupling between the two oscillators vanishes for the same initial condition of two oscillators and the model effectively reduces to a single filtered VdP oscillator. This reduces the problem as parallel evolution of the two oscillators' states separately where the first state variables of both the oscillators represent an ECG-like waveform [13]. In the present work, we optimized the two time delay couplings for each oscillator using a suitable global optimization framework to find out realistic parameters of the above mentioned coupled oscillator system (10). Here, the time delay ($\tau$) and time constant ($T$) values are presented in seconds.

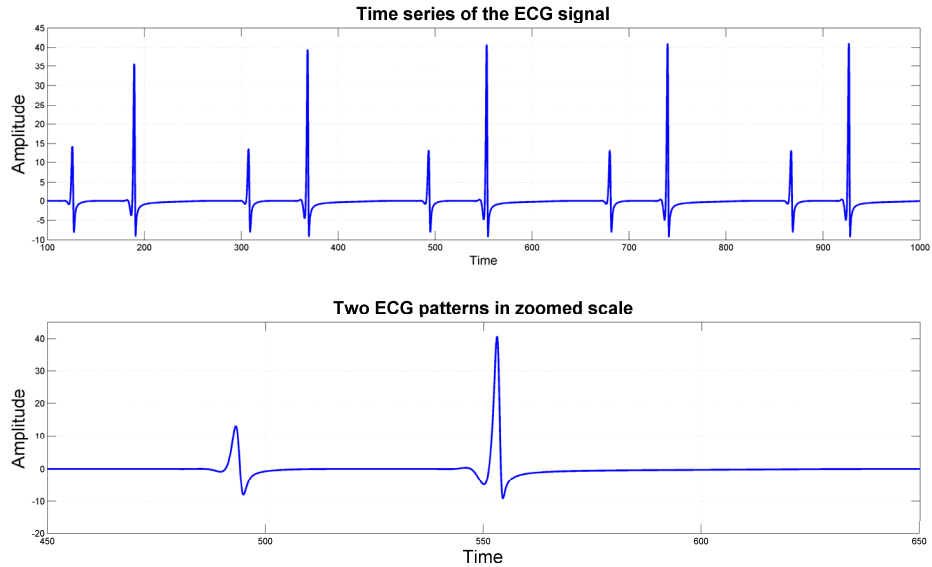

**Figure 1: Presence of two typical ECG like waveforms with the integer order model (10).**

Figure 1 shows time evolution of the first state variable of each filtered VdP oscillators from our simulation study following equation (10), as suggested in [13], which resembles with an ECG waveform. It is evident that there exists two ECG like patterns where the QRS complexes are prominent without any "P" and "T" wave. Since "QRS" wave is an effect of ventricular depolarization, the mathematical model (10) behaves similar to a human heart particularly capturing the ventricular depolarization characteristics. It is also noticeable that among multiple beats, two distinct wave patterns are present in the time series. A zoomed version of the signal with depiction of single beats for each of them has also been shown in Figure 1. The periodic nature of the ECG-like waveforms can be better captured using the phase portrait analysis. In all the figures showing simulated ECG signals from the coupled oscillator model, the x-axes is denoted in terms of the number of samples while the sampling time is 0.01 sec.

In phase space analysis of dynamical systems, plots among different state variables are generally referred. In most cases, these state variables are chosen in such a way that one state becomes differential or integral of the other states and so on. In ECG signal processing literatures, generally the phase space representation between the ECG signals with its delayed versions are studied. Similar studies have also been done involving the first derivative of ECG [48], chaos study in phase portrait [49], phase space reconstruction with time delay [50] etc. Here for simplicity we have chosen the ECG signal obtained from the coupled oscillator model, its derivative and integral as the three axis of the phase space for pictorial representation. Therefore, the other axis of the phase space diagrams is represented by the integrated or differentiated version of the ECG signal after passing it through an integrator and a differentiator ($s^{\pm 1}$ in Laplace domain) respectively. The phase space diagram has been widely used to detect normal and abnormal heart rhythm for example as studied by Roopaei *et al.* [49].

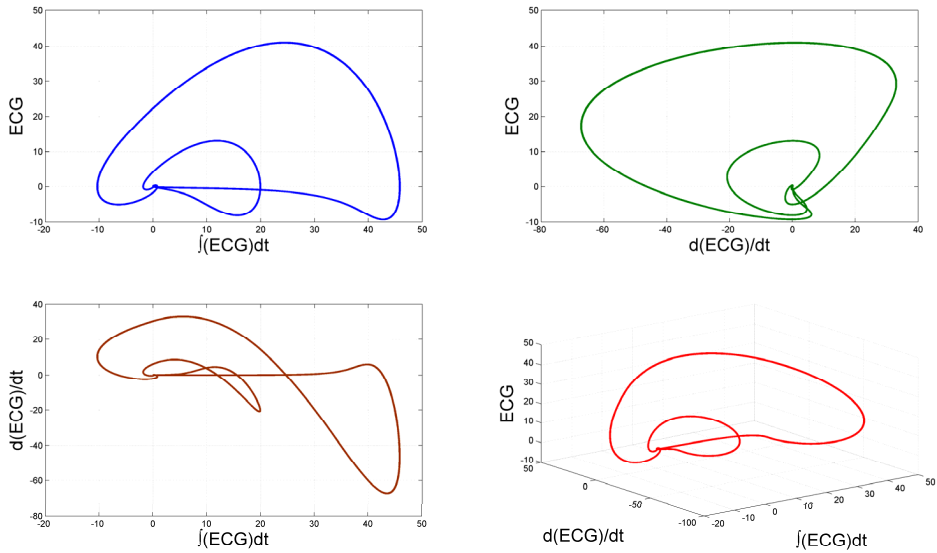

**Figure 2: Phase space depiction of ECG signal obtained from integer order coupled oscillator model (10).**

Figure 2 shows (once again according to our simulations) the phase space diagrams of the ECG waveforms obtained for the coupled oscillator model, given by (10). As mentioned above that two distinct periodic waveforms can be identified from the ECG-like time series with almost similar shape but different amplitudes in Figure 1, which is also captured in Figure 2 as the phase portraits are having two clearly identifiable closed contours which indicate the presence of two periodic ECG wave-patterns. It is obvious that healthy ECG signal should show only one periodic pattern, which is the motivation behind further refinement of the coupled oscillator model over that reported in [13]. It is to be noted that the purpose of the two-coupled oscillator based modelling is to simulate the morphology of an ECG signal. It should not be confused with any anatomical region of the heart. The practical analogy of the coupled VdP oscillator with real structure of human heart is still an open problem and need further exploration.

Regarding the practical analogy of the coupled oscillator model with the human heart it can be summarized that just like any other biological systems, human heart can also be considered as a dynamical system. Any dynamical system evolves through different phases which are known as states, for example generation of P-QRS-T waves are result of different states of periodic electrical polarization and depolarization phenomena in heart. Each state could be characterized by a set of variables evolving over time known as state variables. Therefore, the evolution of the state variables is fundamental in describing the overall temporal dynamics of a system. The traditional way of modelling the evolution of such states is through a set of differential equations involving those state variables. In this paper, we utilized this concept of state variables in a coupled oscillator scenario to explore the possible synthesis of healthy and pathological ECG signals which may shed some light on the actual dynamical phenomena behind the generation of ECG signals.

## 3. Proposed incommensurate fractional dynamical model to generate ECG like waves

Here, we propose a FO incommensurate dynamical model for ECG like waveform generation to show the higher capability of FO coupled oscillator models to describe healthy and unhealthy heart-beats than using the conventional integer order model [13]. The incommensurate FO model of VdP oscillator was first studied in Tavazoei $et\ al.$ [37]. The proposed incommensurate fractional dynamical model for the coupled VdP oscillator system are represented by (11) where orders of the fractional differentiation ($\gamma_1, \gamma_2$) are different for the first two state equations for both of the two identical filtered VdP oscillators. For simulation study, we considered the two time delay terms in the second state variable of the oscillators to be same ($\tau$) to show the effect of FO modelling only, excluding any change in the coupling terms. Similar to the argument, mentioned in section 2.2, the consideration of two time delays being same reduces the model (11) to a single FO filtered VdP oscillator. In model (11), the time constant ($T$) and time delay ($\tau$) are specified in seconds.

$$\left.\begin{aligned}
\frac{d^{\gamma_1} x_1}{dt^{\gamma_1}} &= y_1 + \varepsilon\left(1 - \mu z_1\right) x_1 \\
\frac{d^{\gamma_2} y_1}{dt^{\gamma_2}} &= -x_1 + \alpha\left[x_2(t-\tau) - x_1(t-\tau)\right] \\
\frac{dz_1}{dt} &= \left[\left(\left(|y_1| - y_1\right)/2\right) - z_1\right]/T \\
\frac{d^{\gamma_1} x_2}{dt^{\gamma_1}} &= y_2 + \varepsilon\left(1 - \mu z_2\right) x_2 \\
\frac{d^{\gamma_2} y_2}{dt^{\gamma_2}} &= -x_2 + \alpha\left[x_1(t-\tau) - x_2(t-\tau)\right] \\
\frac{dz_2}{dt} &= \left[\left(\left(|y_2| - y_2\right)/2\right) - z_2\right]/T
\end{aligned}\right\} \quad (11)$$

From extensive simulation study we observed that the periodic pattern of the ECG waves gets lost if the order of the third state equation i.e. the equation contributing the low pass action for magnitude stabilization of the oscillators are changed to take arbitrary fractional value. From our simulation studies we observed that the order of differential equations of the first two state equations may lead to meaningful conclusion regarding indication of healthy and unhealthy ECG like waves using the FO coupled oscillator model given by (11). Since fractional dynamics in the third state equation gives unstable response, similar study using commensurate FO models are not possible, because it does not preserve the ECG like periodic nature of the state variables of the coupled oscillator.

We studied three special cases with the FO dynamical model (11), i.e.

a. Fractional dynamics in the cross-product or nonlinear state equation i.e. $\gamma_1 = \gamma, \gamma_2 = 1$.

b. Fractional dynamics in the time delay coupling state equation i.e. $\gamma_1 = 1, \gamma_2 = \gamma$.

c. Fractional dynamics in the first two state equation i.e. $\gamma_1 = \gamma_2 = \gamma, \quad \gamma \in \Re_+$.

Each of the above class of models yielded different real-life ECG waveforms which may describe the underlying physical process behind generation of such typical electrical waves in human heart. Here a MATLAB/Simulink based model is developed to implement (11) and the Oustaloup's recursive approximation (6) is used to numerically evaluate the fractional derivative of each state variables [19].

Indeed the fractional dynamics is employed in the three state variables of (11) and can be visualized after the states are written in terms of equivalent integral equation form with the memory kernel decaying as a power law instead of a Dirac delta function as in the case of integer order cases. In fact similar dynamical nature can be achieved by very high order IIR filtering of the oscillator outputs which can be compactly represented by only a few number of fractional derivative terms and tuned to match real-world biological signals, like ECG in this case. The high order filtering is automatically employed in the ORA approximation of fractional derivatives as discussed in equation (6).

### *3.1.     ECG like wave generation with different fractional dynamical models*

The three different cases of fractional dynamics being present in the nonlinear term, the time delay coupling term and both the first two state equations are simulated now using the model (11). As expected, the FO coupled filtered VdP oscillator model behaves in a completely different manner as opposed to the integer order model in (10) as shown in Figure 3-Figure 5. The time series generated by varying the fractional derivative operator gives a wide variety of ECG like waves relating various healthy and unhealthy physiological conditions. For all the three cases, mentioned earlier, it is observed that with a higher order ($\gamma$) of fractional differential equation, the difference between two ECG beats are increased which may be an indication of a slow heart rate. Similarly small order of fractional differential equation represents more rapid heart rate. Thus it is clear that the FO models have higher capability of modelling the variation in different heart rates, by virtue of a single parameter i.e. the order of the fractional differential equation. Regarding periodicity of the ECG waves, FO coupled oscillator models are well behaved than the integer order counterpart. In Figure 3-Figure 5, it is observed that most of the ECG beats have equal magnitude, whereas the ECG beats from the integer order model in Figure 1 suffers from the presence of two beats with different amplitude.

Figure 3 shows the ECG like time series using the FO coupled oscillator model (11) by considering the fractional dynamics to appear only in the first state equation, affecting the nonlinear cross-product term (case 11a). It is seen that except differential equation of order $\gamma = 0.7$, rest of the models having $\gamma = 0.8, 1.2, 1.4$ are capable of producing ECG like waveform, at least the "QRS" complex of equal amplitude, excluding the "P" and "T" waves.

Also, for $\gamma = 0.8$, the "Q" and "S" wave amplitudes are almost equal which is rarely seen in real ECG signals.

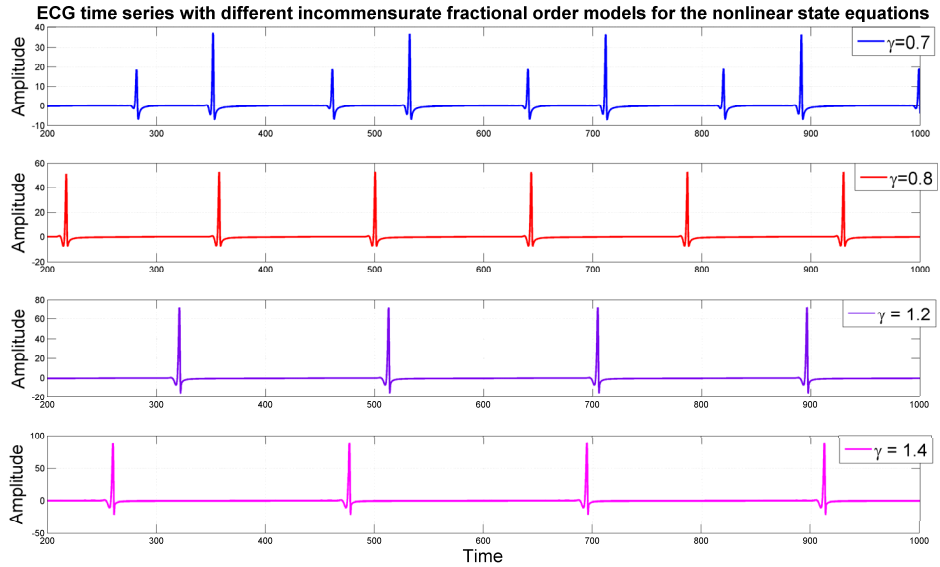

**Figure 3: ECG time series with fractional dynamics in the first state equation (11a).**

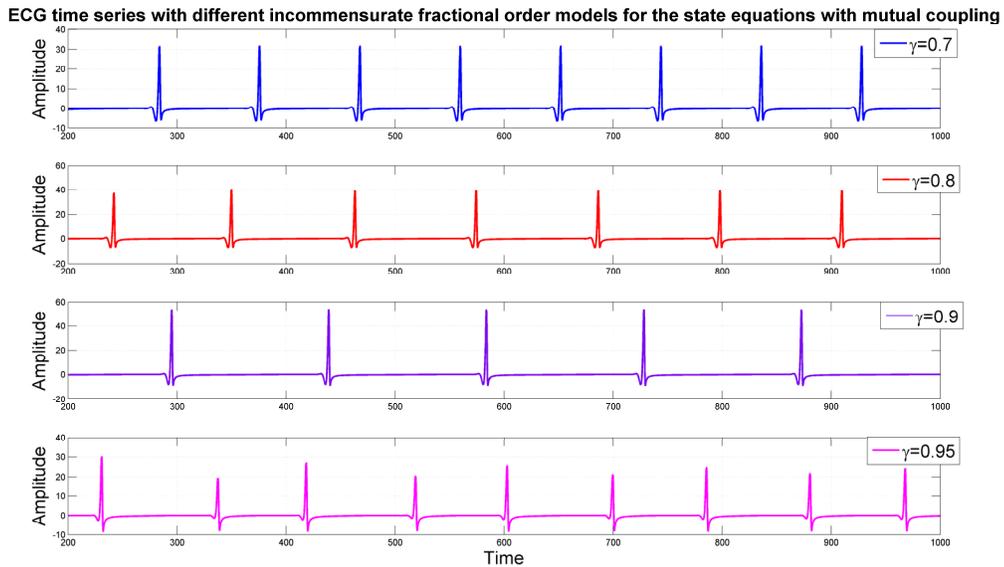

**Figure 4: ECG time series with fractional dynamics in the second state equation (11b).**

Similarly, Figure 4 shows the effect of fractional dynamics being present in the second state equation, affecting the time delay coupling term (case 11b). In this case, it has been observed that fractional differential equation orders $\gamma > 1$ in the coupling equation do not yield periodic nature of the ECG waveform. FO models with $\gamma = 0.8 - 0.9$ produces well-behaved ECG-like waves. For $\gamma = 0.95$ which is closer to the integer order model shows the existence of two different types of QRS complex with two different heights as found in the integer order coupled oscillator model. For the FO oscillator models with $\gamma = 0.7 - 0.9$, the "Q" and "S" wave amplitudes are again found to be almost equal, similar to the previous case.

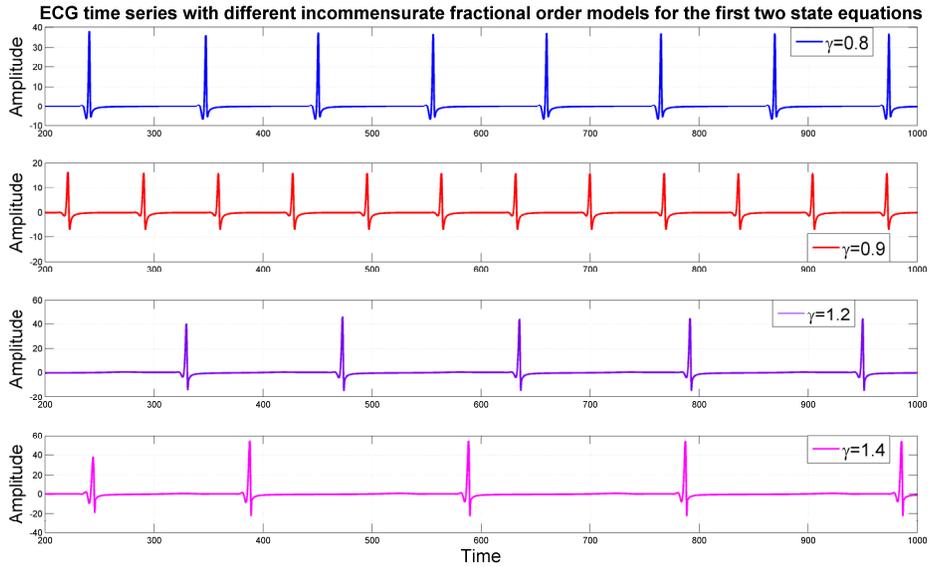

Figure 5: ECG time series with fractional dynamics in the first two state equations (11c).

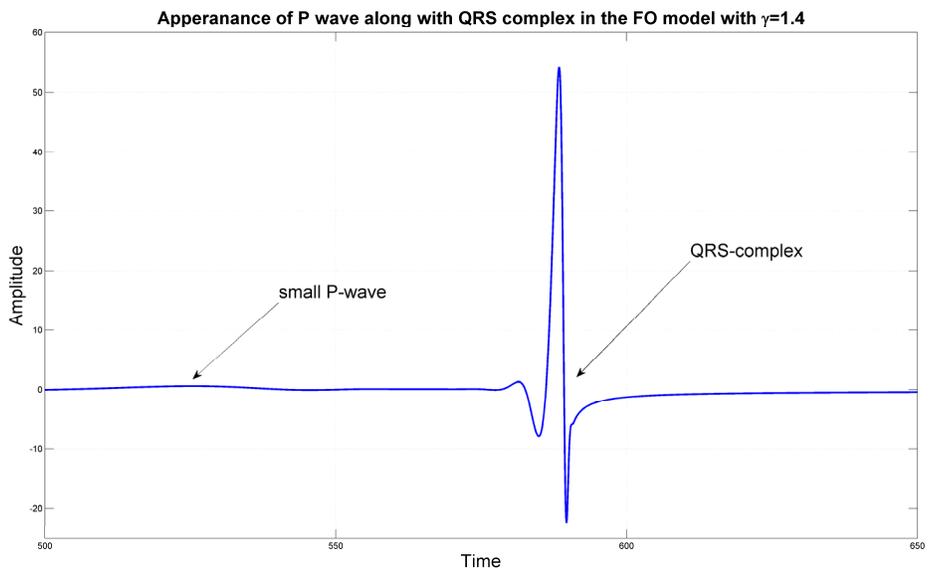

Figure 6: Appearance of P-wave along with QRS complex for the FO oscillator model (11c) with γ=1.4.

With the consideration of same fractional order ($\gamma_1 = \gamma_2 = \gamma$) of the first and second state equations, oscillator models are developed (case 11c) and the resulting waves are shown in Figure 5. For fractional order $\gamma = 0.8$, a particular pattern is observed with the "Q" amplitude being larger than "S" amplitude. The heart rate suddenly increases for the fractional dynamics in first two equations with $\gamma = 0.9$. Also, for $\gamma = 1.2$ a large "S" peak is observed. With $\gamma = 1.4$ we found the best morphology to represent a realistic ECG signal with P-waves along with the QRS complex as depicted in Figure 6. Therefore, amongst different time series obtained from the three cases of the proposed FO coupled oscillator model, the best resemblance with real ECG can be achieved for equal fractional dynamics in the first two state equations with $\gamma = 1.4$. It is arguable that the presence of P-wave in Figure 6 is very small. In a recent study [2], a three coupled oscillator based model successfully reproduced both the P and T waves, under healthy condition only, but its generalization for wide variety of pathological ECG is not known till date and needs further exploration.

### 3.2. Phase space analysis of the FO coupled oscillator model based ECG waves

The phase space diagrams with the ECG waveforms and their integrals and differentials as the three axes have been shown for the three class of FO oscillator models as proposed in equation (11), considering various fractional order of the differential equation below and above one for each state equation. For fractional dynamics being present in the first state variable (11a), it is observed from Figure 7 that in most cases, the phase portraits take circular or elliptic shape e.g. with $\gamma = 0.5, 0.6, 0.9, 1.1$. This essentially implies that the periodic pattern of the ECG waveform collapses to regular oscillations for few of the FO oscillator models which are evident from the limit cycles in Figure 7. For $\gamma = 0.7$, the presence of two periodic waves is also evident similar to that for the integer order model, shown in Figure 1 and Figure 2. For $\gamma = 1.1 - 1.4$, the model with fractional dynamics in the first state equation (11a) faithfully produces ECG like waves, as also evident from Figure 7.

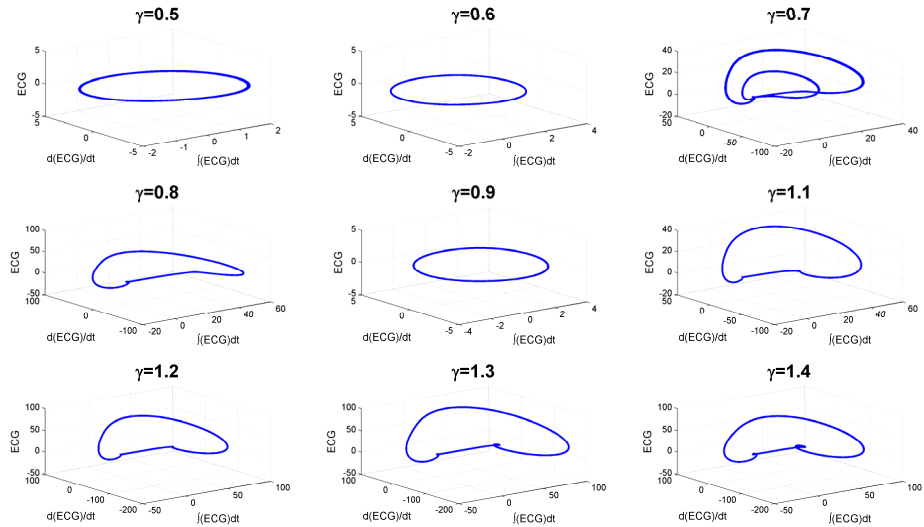

**Figure 7: Phase space representation of oscillator waves with fractional dynamics in first state variable (11a).**

The phase portraits of the oscillator models with fractional dynamics in the second state (11b) are shown in Figure 8. It is seen that only $\gamma = 0.7 - 0.9$ produces the ECG like pattern. Also, instead of regular oscillation, the presence of several limit cycles as a wide ring is observed which is somewhat similar to the chaotic behaviour with random wandering of the states in the phase space. Also, the varying radius of the limit cycles for fractional order higher than one for each state equation indicates oscillations with different amplitude and frequency. The fractional dynamics being present in the first two state equations of the oscillator model (11c) have a much better capability to reproduce ECG like waveforms e.g. $\gamma = 0.8-0.9, 1.1-1.4$. For $\gamma = 0.5-0.8$ normal limit cycles are observed denoting the presence of periodic waveforms similar to normal sinusoids. Also, boundedness of the phase space diagrams of the oscillator models in Figure 7-Figure 9 confirms the stability (including chaotic nature) of the coupled FO oscillator model with the chosen parameters. Analytical stability conditions of linear FO delay differential equations are discussed in [51], [52], but the analytical stability of coupled nonlinear delay differential equation involving fractional derivative is still an open problem and thus we had no other option rather than focussing on the characteristics of the phase portraits.

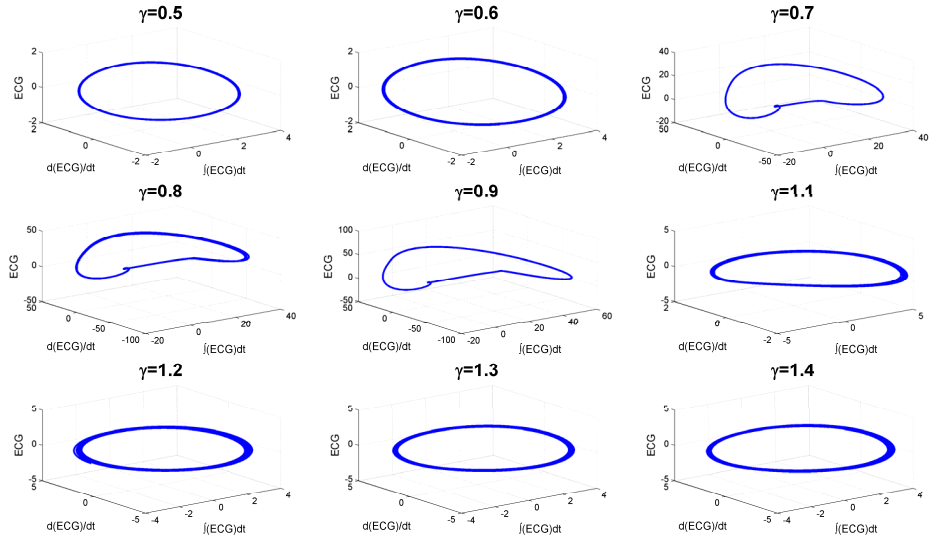

**Figure 8: Phase space representation of oscillator waves with fractional dynamics in second state variable (11b).**

### 3.3. Clinical analogy with healthy and diseased ECG patterns

It has been seen in earlier sections that due to the introduction of the fractional dynamics in various state equations in different combination and the respective values of the fractional orders of the differential equations give rise to ECG waves with high and low heart rate. In ECG literatures, this typical behaviour of very low hear-rate (<60/min) and very high heart-rate (>100/min) are known as *sinus bradycardia* and *sinus tachycardia* respectively [53]. Thus the proposed fractional dynamical modes can easily capture such wide variation in the heart rate with the same generalized template (11) by only suitable modification of the order of the differential equations.

The waveforms generated with the integer order two-oscillator model (10) with equal time delays in coupling is similar to the case of *premature ventricular contraction* where between two normal R peaks a sudden occurrence of distorted "R"-peak can be observed [53]. This typical phenomenon is generally originated from the ventricular muscle. In such a diseased condition generally the "P"-waves are not seen.

The absence of the P wave for the above mentioned models can be compared to the case of *junctional rhythm* of ECG dynamics. In this particular case, often the "P" wave gets superimposed with the "QRS" complex and not seen separately [53]. The source of this particular wave is between the atria and ventricles.

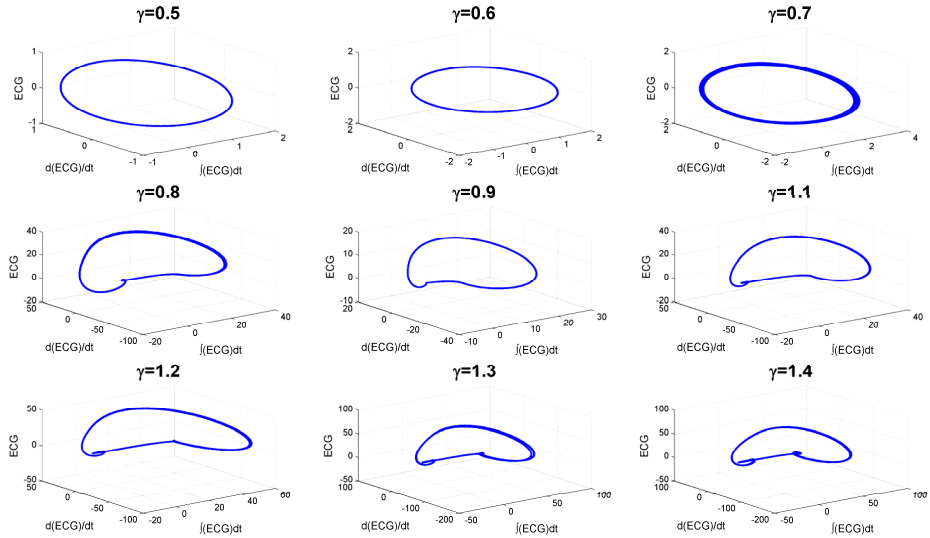

**Figure 9: Phase space representation of oscillator waves with fractional dynamics in first two state variables (11c).**

From the phase space representations it is observed that few of the circular contours are spread in several orbits as such a chaos like random wandering is present in the phase space. This particular phenomenon cannot be easily detected by visual inspection of the ECG time series but the phase space diagrams clearly indicate about such phenomena. Chaotic phase space for ECG signals indicates *ventricular fibrillation*. Such phenomenon occurs with diseased ventricular muscle which generates aperiodic waveform which can be detected by several closed contours in the phase space.

Also, the "S" wave being much taller than the "Q" wave in few FO models in first two state equations with $\gamma = 0.9, 1.2, 1.4$ may be an indication of diseased condition like *ventricular hypertrophy*. This typically occurs due to ventricular overload, stenosis of valves or ventricular septal defect etc. [53]. The above clinical analogies for the proposed model based ECG waveforms establish the generalization capability of the FO dynamical model of coupled filtered VdP oscillator system to produce various ECG like waves under healthy and diseased condition.

The morphological difference between the normal and pathological ECGs considered in this work is summarized in Table 1. Also, conditions of FO models corresponding to each of these cases are highlighted with different mathematical parameters leading to that particular morphology. As discussed previously, the integer order model (10) always generates a beat similar to ventricular premature beat (VPB) even in the case of normal ECG signal. Therefore, it is inadequate in synthesizing real life ECG signals. On the other hand the FO model under different conditions not only reproduces the healthy ECG like signal but also can faithfully generate the morphologies for different pathological conditions. Therefore, FO model can be considered as a more generalized model compared to the integer order counterpart. The present study can be considered as the first attempt of its kind to mathematically model the generation process of healthy and pathological ECG signals with generalized template of fractional order coupled VdP oscillator system. The presented theoretical framework shows that various healthy and diseased ECG signals can be simulated from a generalized model which may be useful in future for characterization of the condition of human heart. For example, by only changing the fractional derivative order, fast or slow heart rate has been reproduced. Also chaotic nature of ECG phase portraits indicating towards arrhythmia has been reproduced with the same generalized model. This article only presents the generalization capability of the model. A complete validation of the model with more real clinical data needs to be explored in future research.

Also, the main goal of the study was to produce ECG like patterns. In ECG signal processing community, it is a common notion to normalize the signal since the scale may be in mV or in ADC unit etc. Also, it is well known the ECG signal amplitudes vary widely in different leads. The output of the coupled oscillators can be scaled up or down using a gain (higher and lower than unity respectively) to match with any ECG scale and hence has not been focussed in the present study.

Table 1: Pathological ECG characteristics and corresponding model parameters

| Pathological cases | ECG characteristics | Model characteristics |
|---|---|---|
| Healthy | clear QRS complex | FO model (11a) with $\gamma=0.8, 1.2, 1.4$; FO model (11b) with $\gamma=0.8, 0.9$; |
| Sinus bradycardia | low heart rate or large R-R interval | FO model (11a) between $\gamma=1.1-1.4$ and with increasing $\gamma$; FO model (11b) between $\gamma=0.7-0.9$ and with increasing $\gamma$; FO model (11c) between $\gamma=1.2-1.4$ and with increasing $\gamma$; |
| Sinus Tachycardia | high heart rate or small R-R interval | FO model (11a) between $\gamma=1.1-1.4$ and with decreasing $\gamma$; FO model (11b) between $\gamma=0.7-0.9$ and with decreasing $\gamma$; FO model (11c) between $\gamma=1.2-1.4$ and with decreasing $\gamma$; |
| Premature ventricular contraction | appearance of distorted QRS complex between two normal beats and absence of P-wave | integer order model (10); FO model (11a) with $\gamma=0.7$; FO model (11b) with $\gamma=0.95$ |
| Junctional | Absence of P wave or superimposed P | All FO models except (11c) $\gamma=1.4$ |

| Rhythm | wave with QRS complex | |
| --- | --- | --- |
| Ventricular fibrillation | Chaotic fluctuation in ECG phase portraits | FO model (11b) with γ=1.2; FO model (11c) with γ=0.7 |
| Ventricular hypertrophy | S wave much taller than Q wave | FO model (11a) with γ=1.2; FO model (11b) with γ=0.95; FO model (11c) with γ=0.9, 1.2, 1.4 |

## 4. Parameter estimation of fractional order coupled VdP oscillator system

### 4.1. Generalization of FO coupled filtered VdP oscillator system with different time delay coupling

It is mentioned earlier that the two oscillators with same initial condition will have similar dynamics. Also due to the fact that the time delay couplings ($\tau$) amongst them being same in (11), the first delayed states for both the oscillators actually cancels each other, thus making the effective coupling signal zero in the second state equations of each oscillator. Also, different initial conditions in the oscillators make the overall coupled system unstable. Thus it is proposed that the time delays of the integer order model should not be same in order to keep the coupling between the oscillators alive. In fact these coupling parameters are very difficult to find out analytically to reproduce an ECG like waveform. Thus we propose a new model structure for the FO coupled filtered VdP oscillator system and also estimated the parameters of it, including the time delay coupling terms for each of the oscillators from a healthy ECG signal. The proposed ECG generation model is thus given by (12).

$$\left.\begin{aligned} \frac{d^{\gamma_1} x_1}{dt^{\gamma_1}} &= y_1 + \varepsilon(1-\mu z_1)x_1 \\ \frac{d^{\gamma_2} y_1}{dt^{\gamma_2}} &= -x_1 + \alpha\left[x_2(t-\tau) - x_1(t-\bar{\tau})\right] \\ \frac{dz_1}{dt} &= \left[\left((|y_1|-y_1)/2\right) - z_1\right]/T \\ \frac{d^{\gamma_1} x_2}{dt^{\gamma_1}} &= y_2 + \varepsilon(1-\mu z_2)x_2 \\ \frac{d^{\gamma_2} y_2}{dt^{\gamma_2}} &= -x_2 + \alpha\left[x_1(t-\tau) - x_2(t-\bar{\tau})\right] \\ \frac{dz_2}{dt} &= \left[\left((|y_2|-y_2)/2\right) - z_2\right]/T \end{aligned}\right\} \quad (12)$$

The fractional differential equations are written in terms of analogous integral equations, for numerical implementation during simulation and to incorporate the initial condition of the state variables as also done in section 2.2.

Here, the fractional dynamics in the first and second state equation has been considered similar to the three classes mentioned in section 3, but only considering the parameters to be unknown. Here in (12) the two time-delay couplings are different i.e. $\tau$ and $\bar{\tau}$. Thus the delayed states corresponding to $x_1$ and $x_2$ becomes different even for the same initial condition. As a result, a finite signal is added to right hand side of the second state equation after getting multiplied by coupling gain $\alpha$, instead of cancelling each other for $\tau = \bar{\tau}$ as in (10) and (11). The idea is now that the parameters of the coupled oscillator (12) can be estimated by minimizing the responses of the coupled oscillator system with a real ECG signal. It is important to note that here only the time delays amongst two coupled systems are considered to be different whereas rest of the parameters like the gains of the mutual coupling ($\alpha$), filter time constants ($T$) and other parameters of VdP oscillator etc. are considered to be same as studied by Kaplan *et al.* [13]. This is due to the fact that optimization based parameter identification with the consideration of all parameters of the coupled system being different may provide better flexibility in ECG signal modelling but would take higher computational resource. Therefore, here we restricted the study for identical incommensurate FO filtered VdP oscillator system, having different time delay couplings ($\tau, \bar{\tau}$) only and not different parameters for each of the oscillators like $\{T, \alpha, \varepsilon, \mu, \gamma_1, \gamma_2\}$.

For all the simulation presented in the paper, the ECG like waves has been generated from the first state of the second oscillator ($x_2$). Since the two oscillators are identical and have equal delay for coupling ($\tau$) and same initial condition in model (11), the time evolution of states $x_1$ and $x_2$ are the same. But for non-identical time delay coupling of model (12) (i.e. $\tau \neq \bar{\tau}$), we considered the second state ($x_2$) which resembles the ECG.

### 4.2. Objective function for optimization based parameter estimation of the FO nonlinear dynamical system

A healthy ECG signal has been taken from the PhysioNet ECG data-bank PTBDB database [54] and the performance index for estimating the coupled oscillator parameters have been defined as (13) relating the real and simulated ECG signals.

$$J = \frac{1}{N} \int_0^T \left| ECG_{real} - ECG_{simulated} \right| dt \qquad (13)$$

For similar parameter estimation of nonlinear FO systems arising in biological applications heuristic optimization algorithms are suggested e.g. in [55], [56]. Here, we used a genetic algorithm for finding the optimum set of oscillator parameters to represent the ECG like waveform. Also, the real ECG signal for estimation has been pre-processed by subtracting the mean or the iso-potential baseline to remove the chance of any possible biased estimation. The time series and phase space representation of the real ECG signal is shown in Figure 10. Since in real ECG signals the baseline often fluctuates and also the P-QRS-T waves exactly do not match for different beats, which appear as different closed contours in the phase space. This happens due to possible introduction of noise in the signal and also due to the inaccuracies of ECG recording instrument. But it unarguable that ECG is an almost periodic signal which must appear as closed contours in the phase space [3].

Equation (13) here denotes the mean absolute error (MAE) of the error signal between the real ECG signal and simulated coupled oscillator based ECG signal. Since both the signals are periodic in nature, even a small difference between them will be integrated over time, thus resulting in a large value of the absolute deviation between two signals. In order to minimize the difference between two periodic sequences, the error signal has been normalized with the number of samples ($N$) to avoid winding up of small but consistent errors. The integral over a finite duration of $T=100$ seconds in the objective function (13) has been implemented numerically using the well-known Trapezoidal rule.

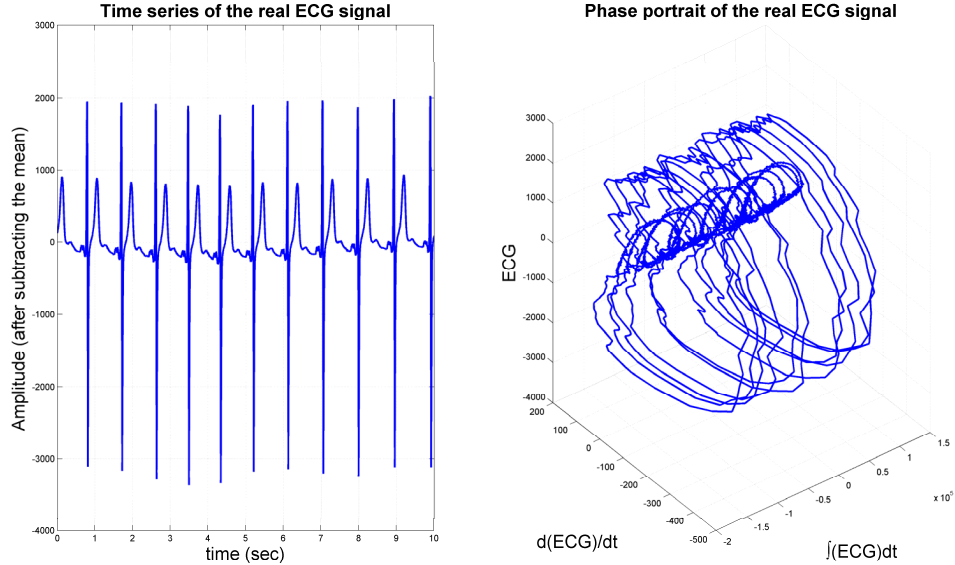

**Figure 10: Time series of a real healthy ECG signal (from PTBDB database of Physionet) and its phase space representation.**

**Table 2: Optimization results for the proposed coupled oscillator systems for producing ECG like waveform**

| Model type | order | $J_{min}$ | $T$ | $\alpha$ | $\varepsilon$ | $\mu$ | $\tau$ | $\bar{\tau}$ | $\gamma$ |
|---|---|---|---|---|---|---|---|---|---|
| Integer order | =1 | 283.7423 | 0.3138 | 0.9902 | 0.7957 | 0.3602 | 0.9303 | 0.5904 | - |
| FO in first state equation | >1 | 283.5751 | 0.8413 | 0.846 | 0.2168 | 0.6985 | 0.3728 | 0.612 | 1.4602 |
| | <1 | 283.69 | 0.9977 | 0.5585 | 0.9085 | 0.426 | 0.0796 | 0.8474 | 0.999 |
| FO in second state equation | >1 | 282.9566 | 0.6143 | 0.4679 | 0.1864 | 0.9882 | 0.3827 | 0.7631 | 1.6903 |
| | <1 | 283.7207 | 0.3247 | 0.3751 | 0.4443 | 0.4305 | 0.6794 | 0.4269 | 0.9067 |
| FO in first and second state equation | >1 | 283.5367 | 0.0932 | 0.8785 | 0.6586 | 0.924 | 0.1588 | 0.7468 | 1.4294 |
| | <1 | 283.7092 | 0.3798 | 0.6729 | 0.9472 | 0.3186 | 0.0141 | 0.6872 | 0.9279 |

The error criterion in (13) may be defined as a squared error instead of absolute error. But for high error the squared term would effectively over-penalize the error which may affect the parameter selection procedure of the oscillator within the optimization. FO model optimization using fractional integral as the performance index has been previously studied by Das *et al.* [57] in the context how the dynamical evolution and memory of the error should be taken into consideration. But in present context, the main goal is to match the ECG signal with the oscillator's output to obtain the optimal set of oscillator parameters, thus we avoided such complex formulation of the objective function.

The objective function (13) has been minimized with the well-known global optimizer known as real coded genetic algorithm (GA). Compared to the gradient based and initial guess based single-agent optimizers like Nelder-Mead Simplex (implemented in *fmincon* or *fminsearch* function of MATLAB) and Simulated Annealing (implemented in *simulannealbnd* function of MATLAB), the GA is less susceptible to get trapped in local minima in the search space [58]. Due to the large population based search and optimization nature of GA, it is widely used in various complex optimization problems where the nature of the landscape is not known *a priori*.

It is well known that genetic algorithm is a stochastic optimization process which can be used to minimize a chosen objective function e.g. here for the minimization of (13). The solution vector which is coded in terms of real variables is initially randomly chosen from the search space which undergoes three stages namely reproduction, crossover and mutation, in each iteration, to give rise to a better population of solution vectors. Reproduction implies that solution vectors with higher fitness values can produce more copies of themselves in the next generation. Crossover refers to information exchange based on probabilistic decisions between solution vectors. In mutation a small randomly selected part of a solution vector is occasionally altered, with a very small probability. This way the solution is refined iteratively until the objective function is minimized below a certain tolerance level or the maximum number of iterations are exceeded. In the present study, the number of population size in GA is chosen as 20. The crossover and mutation fraction are chosen to be 0.8 and 0.2 respectively for minimization of the objective functions (13) to produce optimum set of oscillator parameters.

### *4.3.    Simulation studies with the optimized coupled oscillator parameters*

The optimization is carried out to find out the optimum set of decision variables $\{T, \alpha, \varepsilon, \mu, \tau, \bar{\tau}, \gamma\}$ for the proposed FO coupled oscillator model (12). For similar integer order model $\gamma$ is set to unity and rest of the six variables of the coupled oscillator system are optimized. FO dynamics of higher and lower than unity has been separately investigated since the Oustaloup's recursive approximation for each fractional derivative works only with $\gamma < 1$ [19]. Therefore during simulation of $\gamma > 1$, the orders of the integral equation are split into a traditional integer order (IO) integrator, followed by a FO integrator as suggested by Petras [19]. The optimization results are summarized in Table 2 showing the minima of the objective function ($J_{min}$) indicating the goodness of fit along with corresponding optimum oscillator parameters. From Table 2 it can be observed that the FO models with $\gamma > 1$ consistently give lower error, thus indicating a better representation of the original ECG signal. The time series and phase space representations of the outputs of the optimized coupled oscillator systems are shown in Figure 11 and Figure 12 respectively. It is evident from the simulation results that the model with FO dynamics in the first state equation, affecting the cross-term nonlinearity, is capable of capturing the dynamics of original ECG signal shown in Figure 10. Also, differences in the QRS shape can be observed in different optimized coupled oscillators according to their capability in mimicking the original ECG signal, but the heart rate is nicely captured by all dynamical models having optimum parameters. Parameter estimation of the coupled oscillator system while not considering the two VdP oscillators as identical may give better optimization results but not reported in the present work for the sake of simplicity.

In order to diagnose a diseased heart condition, the system under consideration needs first to be modelled mathematically and then correlate any change in the system parameters with the observation of healthy or unhealthy condition. This paper focuses on mathematical modelling of certain cases of ECG traces and provides a generalized framework to model them from recorded ECG signals using an optimization based approach. In order to establish the connection between the proposed mathematical model and the physiology of heart, further research is necessary. The mapping of spatial localization of different components of the heart and parameters of the mathematical model is still an open problem. The present study shows that by changing only one parameter i.e. the fractional derivative order different physiological conditions can be simulated e.g. increase and decrease in heart rate, appearance of ventricular premature beats and different morphology of QRS complex. This may be an indication that the physical process behind the generation of these typical ECG signals came from the same governing equation which has been attempted to model in this study. The proposed model may have some potential for further exploration with large population of ECG signals with various pathological conditions.

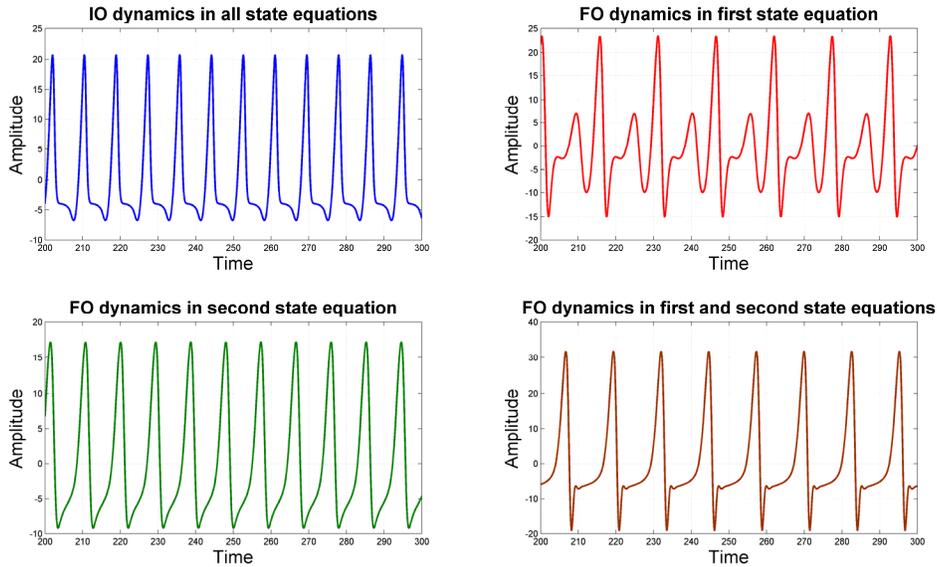

**Figure 11: Time series of ECG like waves with optimized oscillator parameters.**

In fact, there are several evidences of real ECG signals exhibiting significant fractal properties like 1/f noise spectra of QRS complex [59], [60], multifractality [61], [62], fractal correlation of R-R interval [63], long range dependence of ECGs [64], [65], nonstationarity and nonlinearity of ECGs [66], [67] etc. The present work can be considered as the first step towards modelling such wide variety of phenomena using fractional order dynamical systems theory. Since fractional time derivative has an inherent capability of modelling long range correlations of a signal, it is exploited to reproduce some real ECG like patterns under healthy and pathological conditions. But it is still an open problem to reproduce all the above mentioned fractal natures in synthetic ECG and thus further research is needed.

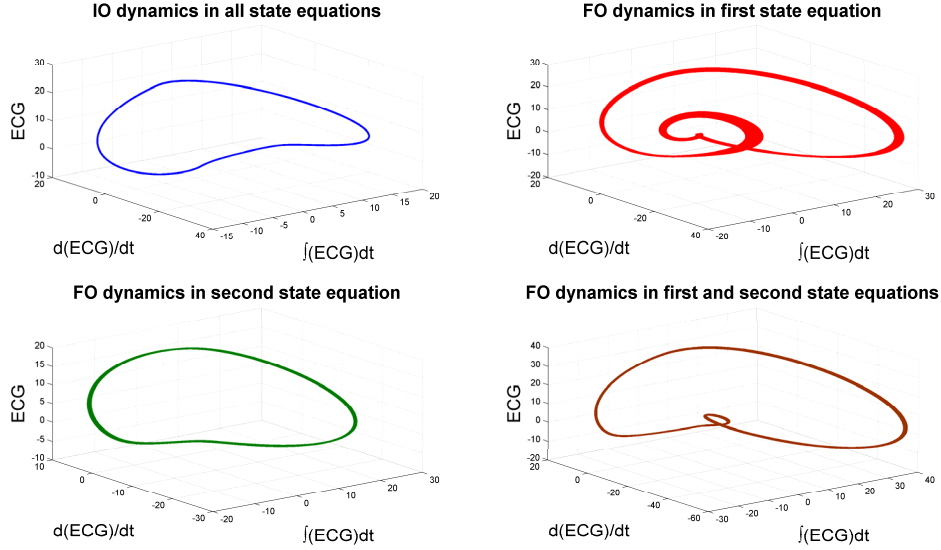

**Figure 12: Phase space representation of ECG like waves with optimized oscillator parameters.**

## 5. Conclusion

A coupled filtered VdP like oscillator system is proposed in this paper having different time delay couplings between them. Different combinations of incorporating fractional dynamics in the different state equations lead to appearance of various healthy and diseased ECG like waveforms with additional control over the heart rate apart from the morphology of the QRS complex. From a real healthy ECG signal, the parameters of the coupled oscillator system are estimated using a global optimization technique. Amongst the first class of FO oscillator models with equal time delay coupling, the fractional order $\gamma = 1.4$ best morphologically matches the ECG waves. The parameter optimization study amongst the second class of FO oscillator models with different time delay coupling gives the best ECG like waves with fractional dynamics being present in the first state equation and with $\gamma = 1.4602$. Future scope of research can be directed towards finding analytical stability of the proposed coupled oscillator system and studying the parametric robustness of the nonlinear model in mimicking ECG like waveforms. Also, the proposed technique may be helpful for person-specific model development and real-time instrument development for model parameter extraction and ECG trace analysis using phase space diagrams in future.

It is understandable that the main goal of analysis of ECG data is to prove doctors with true evidence for their diagnosis of patients. However, we believe that this is only one aspect of the day-to-day clinical practice. The other aspect is to develop understanding about how the heart conditions dynamically evolve over time reflecting different clinical conditions since this is the very fundamental aspect of any cardiovascular disease prognosis. This calls for developing sound mathematical models that may describe the governing dynamics that is physiologically manifested by interaction of heart components (muscular level down to the cell level) under different conditions. If such understanding could be developed then it may be possible to translate that to day-to-day clinical diagnosis with the prerequisite of physically mapping the mathematical dynamics to the physiological components of the heart. If successful, such a method will indeed help in understanding of cardiovascular complicacies and hence improve the diagnostic capability instead of using the currently adopted mostly empirical observation based diagnosis, which in its turn has its own shortcomings. However it is well known that despite several attempts over at least three or four decades still it has not been possible to develop a generalized dynamical model of heart that can faithfully describe the practically observed ECG signals – not to mention the existing inter-person variability. Our attempt was particularly guided by this fact and we showed that introducing fractional order model it is possible to describe the heart dynamics more closely compared to the existing approaches. Indeed the fractional order model has its own physical meaning which may be possible to link with the actual operational philosophy of heart. Therefore our main intention was to mathematically derive the underlying dynamics of heart in a more faithful way rather than making an one-to-one translation of that governing differential equation to physiological components. As a matter of fact such translation is still a completely open question even after several decades of research. Therefore it is conceivable that such dynamical model is still far away from its direct application in day-to-day clinical practice and our model is no exception to that. But given the massive information on human physiology (from organ level down to cell level) becoming available nowadays, thanks to the fast development of computer science and efforts given in several projects, it is also conceivable that in future linking physiology with such fundamental mathematical operation of an organ will be possible and thereby increasing the diagnostic and treatment capability many folds compared to the today's dominant empirical diagnostic methods guided by population statistics that poorly considers person-centric nature of disease symptoms. Our work is the first step in that direction where we developed a novel model for closely approximating the underlying dynamics of heart in the form of ECG.

## Acknowledgement

The work presented in this paper was supported by the E.U. ARTEMIS Joint Undertaking under the Cyclic and person-centric Health management: Integrated appRoach for hOme, mobile and clinical eNvironments – (CHIRON) Project, Grant Agreement # 2009-1-100228. The authors thank the anonymous reviewers for providing constructive comments and useful references which helped to enhance the quality of the paper.